# Coexistence of $Co^{3+}$ and $Co^{2+}$ in ceramic $Co_3TeO_6$; XANES, Magnetization and first principle study

Harishchandra Singh[1], Haranath Ghosh[1], T. V. Chandrasekhar Rao[2], A.K. Sinha[1]

[1]*Indus Synchrotron Utilization Division, Raja Ramanna Centre for Advanced Technology, Indore – 452013, India*

[2]*Technical Physics Division, Bhabha Atomic Research Centre, Bombay, 400085, India*

*Evidence of coexistence of $Co^{3+}$ with $Co^{2+}$ in ceramic $Co_3TeO_6$ through XANES, DC magnetization and first principal studies is provided. XANES along with linear combination fit provide relative concentrations of $Co^{2+}$ and $Co^{3+}$.Temperature dependent DC magnetization exhibits the same antiferromagnetic behavior as observed in single crystal. The presence of both $Co^{2+}$ and $Co^{3+}$ suggests that if the later is in high spin state, the effective magnetic moment is similar to that observed in single crystal studies. In contrast, if $Co^{3+}$ is in low spin state effective magnetic moment is similar to that observed in $Co_3O_4$. It is further shown that both $Co^{2+}$ and $Co^{3+}$ in high spin states constitute a favorable ground state through first principle calculations where Rietveld refined Synchrotron X-ray diffraction data are inputs.*

Multiferroic materials exhibit either coupling between electronic and magnetic orders (type-II) or separate single order parameters (type-I) and have drawn great attention through application as well as fundamental science.[1-4] Cobalt tellurate $Co_3TeO_6$ (CTO) is a newly found multiferroic material which belongs to a novel series of complex metal oxides $A_3TeO_6$ (A=Co, Mn, Ni, Cu) that show interesting physical as well as low temperature magnetic properties, like thermal variations of magnetization, susceptibility, specific heat indicating very complex magnetic structures.[5-6] Neutron diffraction data show that CTO is antiferromagnetic at low temperatures exhibiting subsequent magnetic phase (commensurate to incommensurate and vice versa) transitions.[7-8] Complex spin dynamics (like spin rearrangements/spin flip-flop transitions) along with unusual magnetic structures are also suggested through measurements like magnetization and polarization with temperature and magnetic field.[8] Single crystal CTO crystallizes in monoclinic crystal structure with space group symmetry C2/c, in which $Co^{2+}$ ions occupy five different crystallographic positions, one tetrahedral and four octahedral sites.[7,9] The 3*d* electrons of the Co ions therefore experience different crystal field effects, leading to different electronic/spin configurations. An interesting analogue is $Co_3O_4$ ($Co^{2+}Co_2^{3+}O_4$), which possess spatial inhomogeneity of its magnetic moment via $Co^{2+}$ in tetrahedra and $Co^{3+}$ in octahedral configurations in the normal spinel structure.[10] However, distribution of $Co^{2+}/Co^{3+}$ governed by various external sources like magnetic field, high pressure, temperature etc. have lead to phenomena like diffusive charge transfer, very large effective magnetic moment etc., in the inverse spinel structure.[11-12] Presently, we show for the first time that Co exists both in $Co^{2+}$ and $Co^{3+}$ oxidation states in our ceramic CTO sample. While our first principle simulation favors high spin state of $Co^{3+}$, low spin state of the same cannot be ruled out, accordingly the effective magnetic moment would be, as that in single crystal CTO like or $Co_3O_4$ like.

Structurally, CTO has features similar to those of spinel compounds and hence its low-temperature magnetic properties are likely to be very interesting. For example, the $Co^{3+}$ ions in octahedral co-ordinations prefer in low spin state (S=0). This is because, in octahedral configuration the $t_{2g}$ orbital of Co are lower in energy than that of the $e_g$ and the splitting between them is larger compared to that in tetrahedral configuration. This can lead to large magnetic moment in our sample under high magnetic field. As the strength of Hund's rule coupling and the crystal-field splitting are comparable for $t_{2g}$ and $e_g$ orbitals, the spin state of the cobalt ion would depend, in general, on various factors like doping concentration, crystal structure and external magnetic fields etc.[13] For example, field-induced spin-state transition in $Sr_{1-x}Y_xCoO_{3-\delta}$ and a similar temperature induced transition in $La_{2-x}Sr_xCoO_4$ are linked to low-to-intermediate spin (IS) and/or low-to-high spin (HS) state transitions of $Co^{3+}$.[14] In the present report, we show experimentally that the existence of $Co^{3+}$ along with $Co^{2+}$, and are intrinsic to CTO via X-ray absorption near edge structure (XANES) and high field DC magnetization measurements. The latter, in fact, show that cobalt ions exist in high spin state. Existence of $Co^{3+}$ is further supported by the first principle simulations. Coexistence of weak ferromagnetic (FM) and antiferromagnetic (AFM) interactions at low temperatures, reported earlier,[8,15] may arise due to double exchange interaction via unlike spins ($Co^{2+}$-O-$Co^{3+}$) and super exchange interaction between like spins ($Co^{2+}$-O-$Co^{2+}$ and $Co^{3+}$-O-$Co^{3+}$), depending on their spin state, bond angle etc.[16] DC magnetization studies, carried out under zero field cooled (ZFC) as well as field cooled (FC) conditions, confirm various magnetic transitions reported earlier in a single crystal of CTO.[17] High temperature paramagnetic region of the magnetization data is analyzed through Curie Weiss fit which results in a large value of the effective magnetic moment. This is not expected either from $Co^{3+}$ or $Co^{2+}$ ions separately, indicating mixed valency of Co in our CTO sample.[12]

We would like to emphasize that the finding of $Co^{3+}$ oxidation state is a first step towards experimental verification of some of the theoretical predications made in reference 15. For example, as in single crystals, it is well known that the $Co^{2+}$--O--$Co^{2+}$ network gives rise to AFM coupling between Co ions through super-exchange interaction. Local appearance of $Co^{3+}$ in $Co^{2+}$--O--$Co^{2+}$ network can cause (a) structural distortion, (b) local FM interaction through double-exchange (c) weakening of effective magnetic interaction between the $Co^{2+}$ ions (d) in-commensurate (IC) AFM ordering (For instance, a possible reason for the appearance of the IC-AF order in $Co_3O_4$ is considered to be local structural transitions due to a charge and/or a spin state change of Co ions[10]. We have evidence on some of the above through temperature dependent lattice constant evaluation via low temperature Synchrotron X-ray diffraction, low field magnetization measurements etc. However, we believe that these measurements are meant for providing experimental evidence to some of the theoretical predications made in reference 15. The rest of the letter is organized as follows. Method of synthesis, equipment details and structural characterization of ceramic form of cobalt tellurate $Co_3TeO_6$ are described first. Then we analyze structural and XANES data, leading to a conclusion that CTO contains substantial fraction of Co atoms in $Co^{+3}$ states. An analysis of magnetization data follows, the results of which compliment those of XANES studies. Detailed results of our first principle studies are presented in the first principle simulation subsection followed by conclusions.

Single phase polycrystalline samples of $Co_3TeO_6$ were prepared by conventional solid state reaction in air atmosphere, using commercial $Co_3O_4$ (Alpha Easer 99.7 %) and $TeO_2$ (Alpha Easer 99.99 %). In order to compensate for $TeO_2$ loss at calcination stage, we supplied with additional 10% $TeO_2$ over the stoichiometric quantity. Well ground oxide mixture was first calcined at ~ 700 °C for 12 hrs and then at ~ 800°C for 20 hrs. The resulting sample was pressed into pellets and sintered at ~ 880°C for 4 hrs in a quartz tube furnace. Synthesis under various conditions (air atmosphere vs argon atmosphere) along with detailed characterization using Synchrotron X-ray diffraction (SXRD), Energy dispersive spectroscopy (EDS), Scanning electron microscopy (SEM), X-ray photo electron spectroscopy (XPS) and XANES for Co pre-edge, Te $L_3$- edge etc. will be reported elsewhere.[18] XANES and SXRD measurements were performed on angle dispersive X-ray diffraction (ADXRD) beamline (BL-12) at Indus-2 synchrotron source.[19] The beamline consists of a Si (111) based double crystal monochromator and two experimental stations, namely, a six circle diffractometer with a scintillation point detector and image plate based Mar 345 dtb area detector. The latter was used for present SXRD measurements and image plate data was processed using Fit2D program.[20] Photon energy for SXRD was calibrated by using SXRD pattern of $LaB_6$ NIST standard. XANES measurements were carried out in transmission mode at room temperature. Sample absorption around Co K-edge (7.709 keV) was measured by monitoring incident and transmitted intensity with the help of two ionization chambers placed before and after the sample respectively. Photon energies below and above the Co K-edge were resolved using Si (111) double crystal monochromator and calibrated with Co metal. Energy resolution ($\Delta E/E$) was estimated to be $1.5 \times 10^{-4}$ around the Co K-edge. Reproducibility of energy scale was estimated to be within 0.05 eV. XANES spectra were recorded in steps of 1 eV. DC magnetization was measured in the temperature region 5K-300K using a SQUID magnetometer (M/s. Quantum Design, model MPMS), under both zero field cooled (ZFC) and field cooled (FC) conditions. Sample was taken in a clear gelatin capsule, which contributes insignificant and temperature independent diamagnetic moment.

The Synchrotron X-ray diffraction pattern revealed that CTO stabilized in the monoclinic symmetry (space group: C2/c) akin to CTO single crystal.[6-9] The Rietveld refinements analysis on the SXRD data was done using the FULLPROF program.[21] The simulated pattern is shown in Fig. 1 along with experimental data. The bottom blue curve corresponds to the difference between the observed and calculated diffraction patterns. Satisfactory matching of the experimental data with the calculated profile of the SXRD pattern and the corresponding reliability factors (noted in Fig. 1) confirm that the fit obtained is reasonably accurate. The extracted lattice parameters **a = 14.8061(5)Å, b= 8.8406(3)Å, c= 10.3455(4)Å, β=94.819(2)°**, show marginal differences over those of CTO single crystal and earlier reported ceramic samples[5-9]. One of the important differences in our study from earlier reported crystal structure lies in the typical bond distances of Co-O and Te-O. The Co-O bond lengths (obtained from the Rietveld refined structure) vary from 1.84Å to 2.87Å in the present study as compared to 1.97Å to 2.93Å reported earlier[7, 9, 17]. In addition, the typical bond distance of Te-O in our case ranges from 1.69Å to 2.28Å as compared to that reported earlier from 1.88Å to 1.98Å, in slightly more distorted $TeO_6$ octahedra. It may also be possible that some of the Co sites will have $Co^{3+}$ instead of $Co^{2+}$ which would further give rise to contraction of CoO6 octahedra and/or CoO4 tetrahedra.[22] This might result in the shortening of the Co-O bond lengths accompanied by an increase in Te-O bond lengths at least for that oxygen which are connected to both Co and Te. Further details of the microstructure variations in the synthesized CTO will be included elsewhere.[18]

Observed micro-structural variations are possibly due to the observation of $Co^{3+}$ (to be discussed in next section) in addition to $Co^{2+}$ for ceramic CTO may be due to various reasons like oxygen excess[23], cations vacancy[24] etc. However, our XANES results on Te $L_3$ edge confirmed the +VI state of Te in CTO, may indicate a possibility of Co/Te vacancy in addition to that of oxygen reduced atmosphere (Argon atmosphere) synthesis, which has been covered elsewhere.[18] Elemental analysis using energy dispersive spectroscopy (EDS) at two spot shows the atomic compositions are Co2.97(4)/Te1.03(3) and Co2.99(4)/Te1.01(3), respectively if the sum of the cations is assumed to be 4. The oxygen content determined with the same for the ceramic CTO samples was determined as 5.98(3) and 5.99(3). Visualization of CTO structure, polyhedral distortion and bond length analysis has been done using VESTA.[25]

XANES is an element specific spectroscopic tool which provides important information to the oxidation states, local coordination and hybridization effect of orbitals of the elements present in the sample.[26] XANES measurements were performed to check the oxidation state of Co in CTO. Fig. 2 shows edge step normalized XANES spectra of CTO along with Co metal foil (for energy calibration) and cobalt oxide standards. XANES spectrum is divided into four parts, namely, the pre edge, the main edge, white line and the extended edge region.[27]Among the four features, the main edge (rising edge, Fig. 2) is attributed to 1s→4s monopole transition. This rising edge gives information about the oxidation state of the absorbing atom in the sample, as main edge position shifts to higher energy due to core hole shielding effects.[27] We emphasize here only on main edge part of the XANES spectra. From Fig. 2, it is clear that Co in CTO sample cannot be in purely $Co^{2+}$ state. The main edge is shifted toward higher energy side from that of the CoO standard. To understand the shift better, we carried out a detailed analysis of XANES data. For the analysis, we have used XANES spectra of three standard samples, viz, Co metal, CoO and $CoF_3$ of known oxidation states of Co (0, 2 and 3 respectively), along with CTO sample. The spectra clearly show that the main edge position of CTO sample lies between that of CoO and $CoF_3$. This observation indicates that Co in CTO is not in entirely $Co^{2+}$ state, in contrast to earlier reported results on single crystal and ceramic samples[5-9, 17] and suggests the presence of $Co^{3+}$ in addition to $Co^{2+}$. It is possible to calculate the concentration[27] of $Co^{2+}$ and $Co^{3+}$ in CTO by noting their energy positions and using a simple linear combination formula, "Energy positions of CTO sample = {Energy position of CoO × $x$ + Energy position of $CoF_3$ × (1-$x$)} /100", where $x$ is the calculated concentration of $Co^{2+}$. Energy position of main edge in XANES spectra of a sample may be determined either as the energy corresponding to ~0.5 absorption or maximum energy value of a first order differentiated spectrum.[27] Above formula has been derived by assuming a linear dependence of the chemical shift on the average valence by which one can obtain quantitative information on the valence state of Co in CTO.[28] Based on the above idea, quantitative phase composition analysis by Linear Combination Fitting (LCF) method on XANES data, using software Athena, has been done.[29] LCF fit gives $Co^{2+}/Co^{3+}$ in CTO to be ~ 60/40. Quantitative approach used in XANES experiments is based on statistical goodness-of-fit and may not assure the accuracy of the components obtained from the LCF fittings.[30] However, in our recent work, we have found that the concentration of two phases of cobalt oxide estimated using LCF technique matches well with that obtained from Rietveld refinements on SXRD data.[27] LCF works on least square fit algorithm and fits the unknown sample as a linear combination of standards.[29] For present study, CoO and $CoF_3$ XANES spectra are used as standards. The experimental data and the best fit obtained are presented in Fig. 3. The goodness-of-fit is judged by R (residual) and chi-square values.

It was reported earlier that Co exists in various spin states, *viz.*, low, intermediate and high spin. $Co^{2+}$ exist mainly either in low or high spin states, whereas $Co^{3+}$ may exist in all the three above stated states. Thus it is very important to find out which spin state of $Co^{3+}$ is favorable in our sample. For this we have carried out first principal *ab initio* simulations for total energy calculations considering all the possible spin configurations of $Co^{2+}$ and $Co^{3+}$. We shall discuss details of these calculations in the later section. These results, however, have further been found consistent with the observed large effective magnetic moments of 8.56 $\mu_B$ per formula unit (f. u.) obtained using high field dc magnetization measurements detailed below.

DC magnetization of CTO has been recorded as a function of temperature under an applied field of 10 kOe. Fig. 4 shows the measured magnetization under zero field cooled (ZFC) and field cooled (FC) conditions, and the magnetic behavior is similar to that observed for a single crystal, including two main magnetic transitions around 29K ($T_{N1}$) and 18K ($T_{N2}$).[17] Besides, faint branching between the ZFC and FC magnetization curves has been observed at very low temperatures. We have fitted the high temperature part of the dc magnetization data to Curie-Weiss law (equation is shown in the Fig. 4) to check the nature of magnetic interactions and to estimates the corresponding effective magnetic moment. Negative value of Θ obtained here indicates presence of antiferromagnetic interactions, the same as that reported earlier.[17] *Effective magnetic moment estimated from the fitted Curie-Weiss parameters is large (8.65$\mu_B$/f.u.) as compared to that obtained in case of a single crystal*[17](Given the fact that the $Co^{3+}$ prefers octahedral sites with low spin.). In most of the literature reports,[6-9, 17] the effective magnetic moment of $Co^{2+}$ is noted to be ~ 4.73$\mu_B$ per Co ion in CTO (orbital plus spin contribution), assuming high spin state of $Co^{2+}$. The same value in our case is 8.56 $\mu_B$ per f.u., higher than the previous one. On the other hand, spin only values of effective magnetic moment for $Co^{3+}$ are 0, 2.83 and 4.93 $\mu_B$ for low, intermediate and high spin states respectively.[31] Hence we conjecture that Co in ceramic CTO exists as both $Co^{2+}$ or $Co^{3+}$ and the net effective magnetic moment is a weighted average of their individual contributions. Therefore, the spin state of Co is also very crucial for this case. To have some clue about the same, first principle simulation has been performed (discussed latter).

The significance of mixed valance state of Co may be understood by considering its analogy with $Co_3O_4$. In case of normal spinel, antiferromagnetism arises only due to $Co^{2+}$, as $Co^{3+}$ generally favors a low spin state. However, large effective magnetic moments (~ 60 % larger than normal spinel case) are observed in inverse spinel structure, corresponding either to spin state transition from low spin to high spin of $Co^{3+}$ or diffusion of $Co^{3+}$ into $Co^{2+}$site and/or vice-versa.[12] We have a similar scenario of $Co^{2+}$ and $Co^{3+}$ in CTO with high spin and low spin state respectively. If either $Co^{3+}$ with low spin diffuses to $Co^{2+}$ state or aligns into a high spin state due to high magnetic field, the observed large effective magnetic moment in our case can also be explained as in inverse spinel $Co_3O_4$ case.[12] As observed in the XANES and LCF results, the fraction of $Co^{3+}$ in CTO is around 40%. Hence the following formula to estimate such large effective magnetic moment may be used,[12]

$$\mu_{eff} = \mu_B(Co^{2+}) + \mu_B(Co^{3+}) = \mu_B(Co^{2+}) + \sqrt{(4 \times S(S+1) + L(L+1))}$$

Where S (= 2*x*: *x* = 0.4, the fraction of Co atoms with $Co^{3+}$ valence) and L are the spin and orbital quantum numbers respectively. Therefore, such large effective magnetic moment is also definitive indication of mixed charge state of Co. It may also be possible that some of the $Co^{3+}$ are stabilized at tetrahedral sites out of existing $Co^{3+}$ (40%) in the high spin state configuration, leading to higher magnetic moment .[10-12] In the next subsection we shall see that large spin states of $Co^{2+}$and $Co^{3+}$ is the most favorable one.

The first principle *ab-initio* simulations are performed using Material Studio 6.1 CASTEP package[32] that employs density functional theory using plane-wave pseudo-potential method. The first principle plane wave basis set calculations are performed within the generalized gradient approximation (GGA) using Perdew-Burke-Enzerhof (PBE) functional.[33] Ultra-soft-pseudo-potentials, plane wave basis set with energy cut off of 380 eV and SCF tolerance $10^{-6}$ eV/atom are used. Brillouin zone is sampled in the k space within Monkhorst-Pack scheme and grid size for SCF calculation is (2×4×2). We have performed first principle *ab-initio* simulations, in which the lattice parameters *a, b, c, β* obtained from the Rietveld refined room temperature SXRD data on CTO samples in C2/c phases are used as fixed inputs. CTO has 120 atoms in its unit cell, containing 36 Cobalt, 72 Oxygen and 12 Te atoms. Thus it is challenging to perform proper first principle *ab initio* theory of such a material. Cobalt atoms in CTO occupy five different crystallographic sites, one ($Co_1$) at 4*e* and remaining four ($Co_2$, $Co_3$, $Co_4$ and $Co_5$) at 8*f*.[7-9, 17] Out of these, $Co_5$ exists in tetrahedral configuration while others are in octahedral configuration. The purpose of our simulations is to obtain a better understanding of the ground state configurations that justify the observed XANES and magnetization data. For example, whether coexistence of $Co^{2+}$ states with certain percentage of $Co^{3+}$ (namely ~ 40 to 45 % as emerges from XANES studies) is energetically favorable or not. In order to ascertain the same, we have carried out single point total energy calculations using density functional theory based CASTEP module of material studio 6.1. The results of total energy calculations for all possible arrangements of $Co^{2+}$ and $Co^{3+}$ in terms of their spin states and magnetic alignments are carried out; only some typical results have been tabulated in table 1. The second column of table 1 tabulates the nature of initial spin configuration between $Co^{2+}$ (and/or $Co^{3+}$) ions via oxygen. The spin arrangement between $Co^{2+}$ and $Co^{3+}$ is also kept in AFM configuration. For 33% of $Co^{3+}$ in intermediate spin, a FM initial spin arrangement with $Co^{2+}$ (high spin) resulted in unfavorable total energy (+ 87.44 meV), not included in the table. The third and fourth columns represent the crystallographic positions of $Co^{2+}$ and $Co^{3+}$ in CTO structure and their relative concentrations whereas the fifth and sixth columns indicate their spin states. Finally, the single point total energy is shown in the seventh column respectively.

Based on the earlier reports[6-9, 17] on CTO single crystal, we have taken antiferromagnetic arrangement first. One may note that the configurations in which $Co^{2+}$ is in high spin state in combination with high spin of $Co^{3+}$ have favorable magnetic ground states. However, low/intermediate spin state of $Co^{3+}$ is also a possible magnetic ground state. The total energy calculations presented in table 1 depend on the fraction of $Co^{3+}$ present and its neighboring spin configurations etc. As tabulated, $Co^{2+}$ is in high spin state and is antiferromagnetically coupled via oxygen to another $Co^{2+}$. As per Goodenough-Kanamori-Anderson (GKA) rule,[16] some of the arrangements are also aligned ferromagnetically (FM) depending on the angle $Co^{2+}$–O – $Co^{2+}$ (above or below 120). The configurations in which both $Co^{2+}$ and $Co^{3+}$ are in high spin states (Sr. No. 4, 7, 10) have the most favorable total energy.

We have synthesized ceramic form of cobalt tellurate $Co_3TeO_6$ (CTO) by conventional solid state reaction route using *off-stoichiometric* ratio of $Co_3O_4$ and $TeO_2$ to form single phase CTO. We have structurally characterized CTO samples using Synchrotron X-ray diffraction (SXRD) and have determined the space group and lattice parameters. Noticeable differences in the Co-O bond distances are seen between our sample and others reported in literature. A single crystal sample, for example, was reported to show Co-O bond distances of 1.97Å to 2.93Å, while that in our case is 1.83Å to 2.87Å. Such shorter bond lengths in our sample may possibly be due to existence of $Co^{3+}$ oxidation state of Co. Using XANES measurements, we estimate the concentrations of $Co^{2+}$ and $Co^{3+}$ to be around 60 and 40 % respectively. High field dc magnetization measurements indicate magnetic transitions at ~18K and ~29K consistent with those observed in a single crystal. However, a large effective magnetic moment ~8.5 $\mu_B$ per f. u. obtained by us via Curie Weise fit of the high temperature part of the dc magnetization data is not reported so far. We have provided a possible scheme to explain such a large magnetic moment through existence of both $Co^{2+}$ and $Co^{3+}$. First principle total energy calculations using CASTEP show that existence of $Co^{3+}$ may be energetically favorable in ceramic CTO. Therefore, we conclude from our XANES, magnetic measurements as well as first principle simulations that $Co^{3+}$ is intrinsic to ceramic CTO. This is to our knowledge a pioneering report. We believe that the ceramic CTO with so large magnetic moment will lead to further fundamental and application oriented research.

Authors are grateful to Dr. S. K. Deb. Dr. G. S. Lodha for his support, encouragement and M. N. Singh for his valuable helps in XANES measurements. We thank Pragya Tiwari for elemental energy dispersive spectroscopy (EDS) measurements. Mr. Harishchandra Singh wishes to acknowledge Homi Bhabha National Institute, India for providing research fellowship during his Ph. D work.

## Table and Figure captions

**Table 1:** Total energies of all possible antiferromagnetic and ferromagnetic spin configurations for the coexistence of $Co^{2+}$ and $Co^{3+}$ in $Co_3TeO_6$. The energies have been tabulated with respect to the energy of $Co^{2+}$-O-$Co^{2+}$ antiferromagnetic spin arrangement ($Co^{2+}$ as 100%) as reference. The lowest-energy spin arrangement is given by $Co^{2+}$-O-$Co^{2+}$: AFM+GKA as shown in the table for the respective concentration of $Co^{3+}$.

**Figure 1:** Synchrotron powder X-Ray diffraction patterns recorded at room temperature for ceramic CTO sample. Open red circles represent the raw data, solid black line is the fit obtained by the Rietveld refinements using the monoclinic structure with C*2/c* space group. The vertical bar lines show Bragg reflections. Zigzag line beneath the pattern shows the difference between the observed and calculated Intensity. SXRD pattern have been taken using photon energy of wavelength $\lambda = 0.94805(2)$Å.

**Figure 2:** Edge step normalized Co-K edge XANES spectra of ceramic CTO sample. Co metal is used for photon energy calibration. Cobalt oxide (CoO and $CoF_3$) standards are shown together to obtain Co valency in CTO.

**Figure 3:** LCF fit of CTO raw data with CoO and $CoF_3$ standard samples and linear dependence of oxidation states as a function of energy obtained from derived formula, as in text (inset).

**Figure 4:** High magnetic field DC magnetization in ZFC/FC protocol indicates the same antiferromagnetic behavior as that of a single crystal and Curie-Weiss fit of the same dc magnetization in FC mode (inset). $\mu_{eff}$ has been calculated through $\boldsymbol{\mu_{eff} = \sqrt{(8C)}\ \mu_B}$ per f. u.

**Table 1**

| Sr. No. | Configuration | $Co^{2+}$ (%) | $Co^{3+}$ (%) | Spin state $Co^{2+}$ | Spin state $Co^{3+}$ | Total relative Energy (meV) |
|---|---|---|---|---|---|---|
| 1. | AFM $Co^{2+}$-O-$Co^{2+}$ | 100 ($Co_1$,$Co_2$,$Co_3$, $Co_4$, $Co_5$) | 0.0 | High | -- | 0 |
| 2. | AFM $Co^{2+}$-O-$Co^{2+}$ | 78 ($Co_1$,$Co_2$,$Co_3$,$Co_4$) | 22 ($Co_5$) | High | Interme-diate | -267.16 |
| 3. | AFM $Co^{2+}$-O-$Co^{2+}$ | 78 ($Co_1$,$Co_2$,$Co_3$,$Co_4$) | 22 ($Co_5$) | High | Low | -316.31 |
| 4. | AFM $Co^{2+}$-O-$Co^{2+}$ | 78 ($Co_1$,$Co_2$,$Co_3$,$Co_4$) | 22 ($Co_5$) | High | High | **-305.18** |
| 5. | $Co^{2+}$-O-$Co^{2+}$ AFM+GKA | 67 ($Co_2$,$Co_3$$Co_4$) | 33 ($Co_5$,$Co_1$) | High | Interme-diate | -381.51 |
| 6. | $Co^{2+}$-O-$Co^{2+}$ AFM+GKA | 67 ($Co_2$,$Co_3$$Co_4$) | 33 ($Co_5$,$Co_1$) | High | Low | -341.49 |
| 7. | $Co^{2+}$-O-$Co^{2+}$ AFM+GKA | 67 ($Co_2$,$Co_3$$Co_4$) | 33 ($Co_5$,$Co_1$) | High | High | **-433.00** |
| 8. | $Co^{2+}$-O-$Co^{2+}$ AFM+GKA | 58 ($Co_1$,$Co_4$,Co) | 42 ($Co_2$,$Co_3$) | High | Interme-diate | -117.19 |
| 9. | $Co^{2+}$-O-$Co^{2+}$ | 58 | 42 | High | Low | -450.32 |
| | AFM+GKA | ($Co_1$,$Co_4$,$Co_5$ | ($Co_2$,$Co_3$) | | | |
| 10. | $Co^{2+}$-O-$Co^{2+}$ | 58 | 42 | High | High | **-529.89** |
| | AFM+GKA | ($Co_1$,$Co_4$,$Co_5$) | ($Co_2$,$Co_3$) | | | |

**Fig. 1**

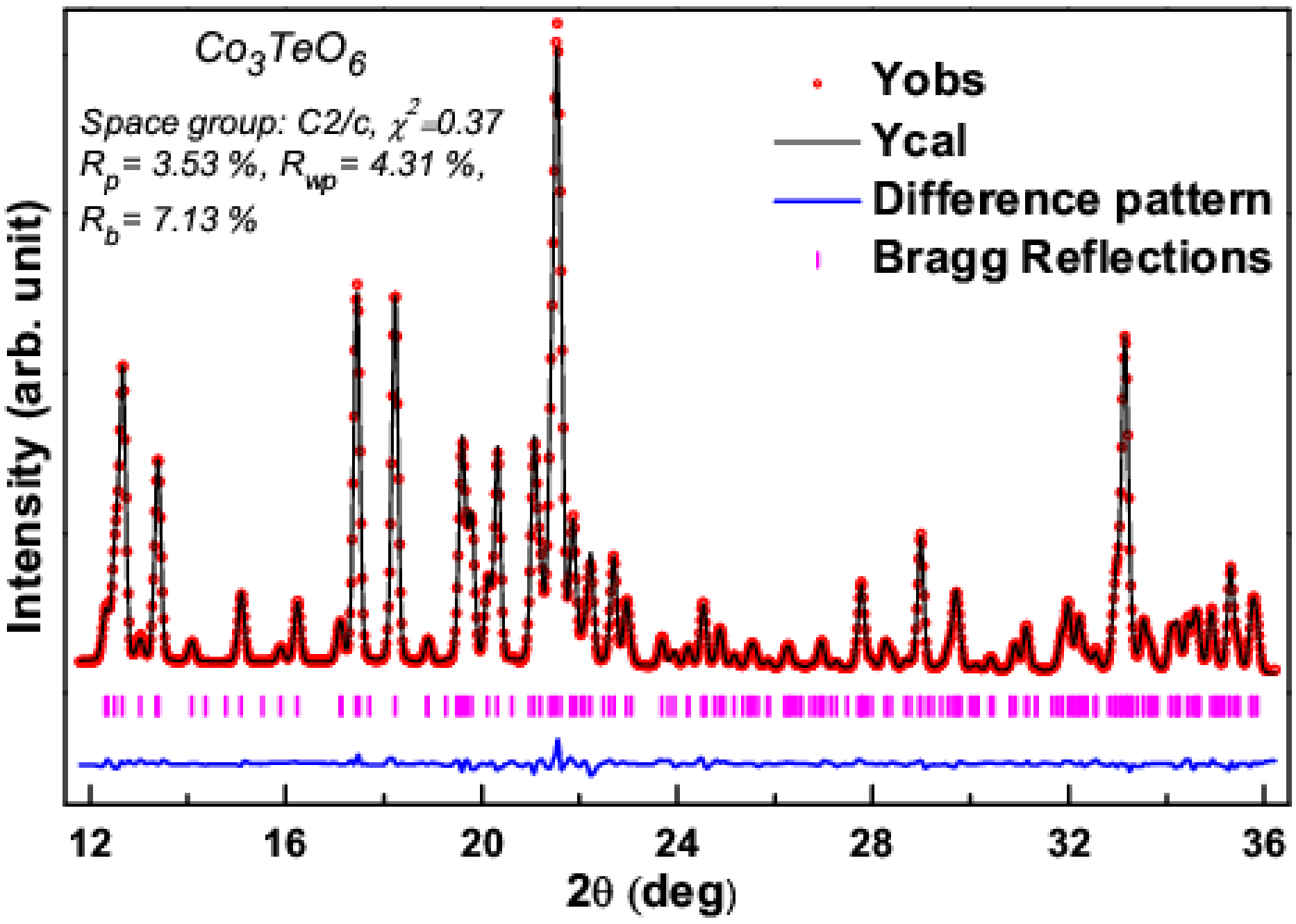

**Fig.2**

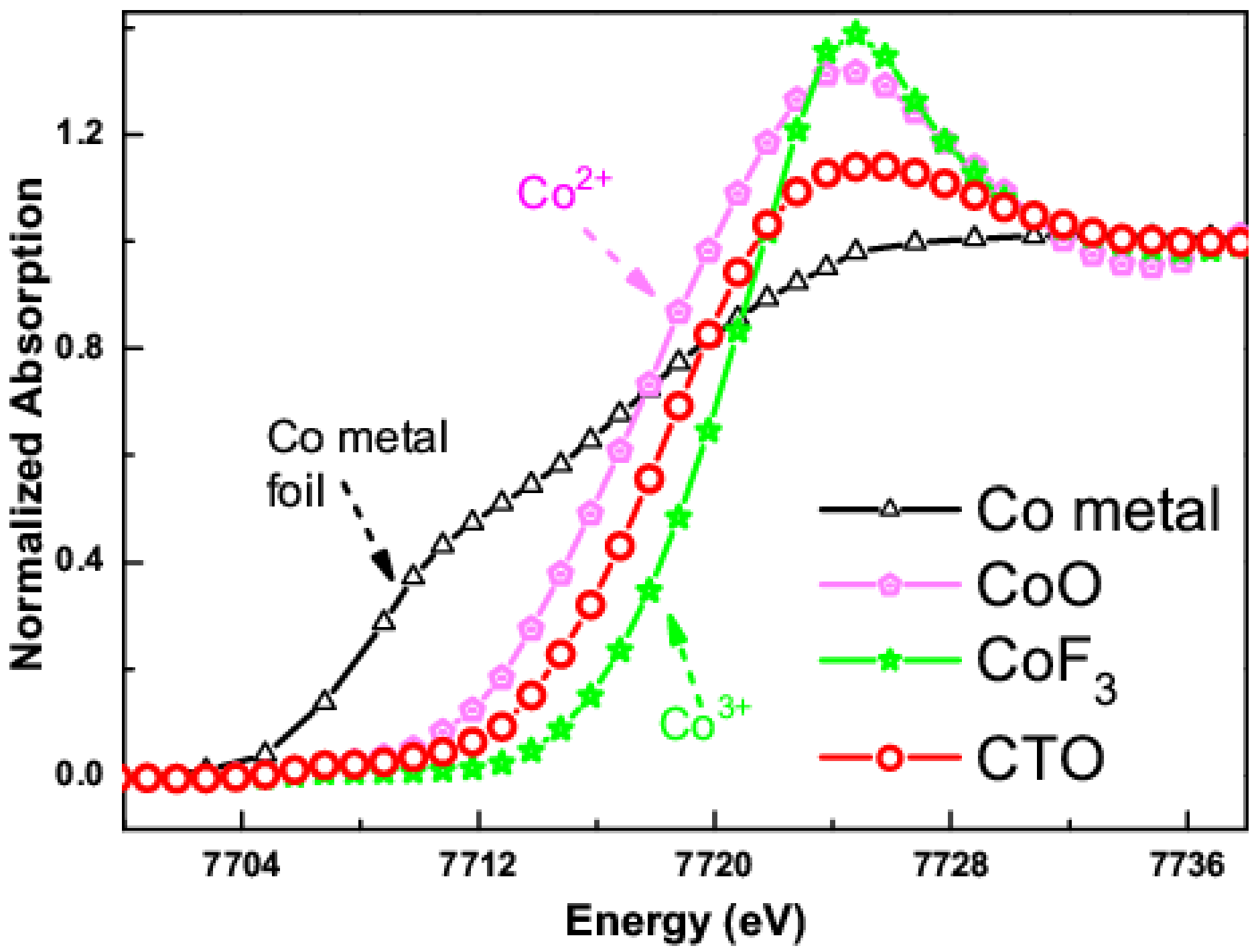

**Fig.3**

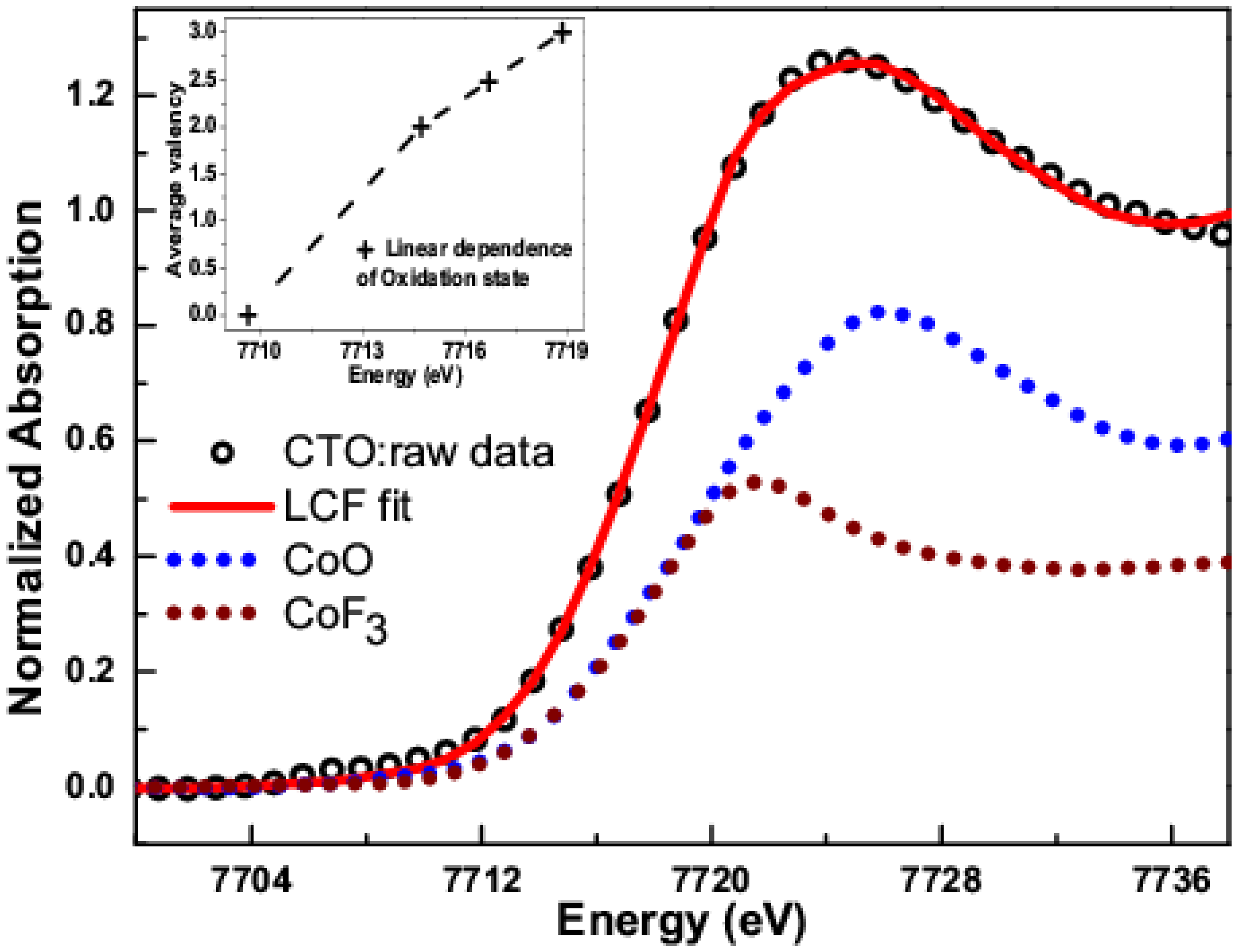

**Fig.4**

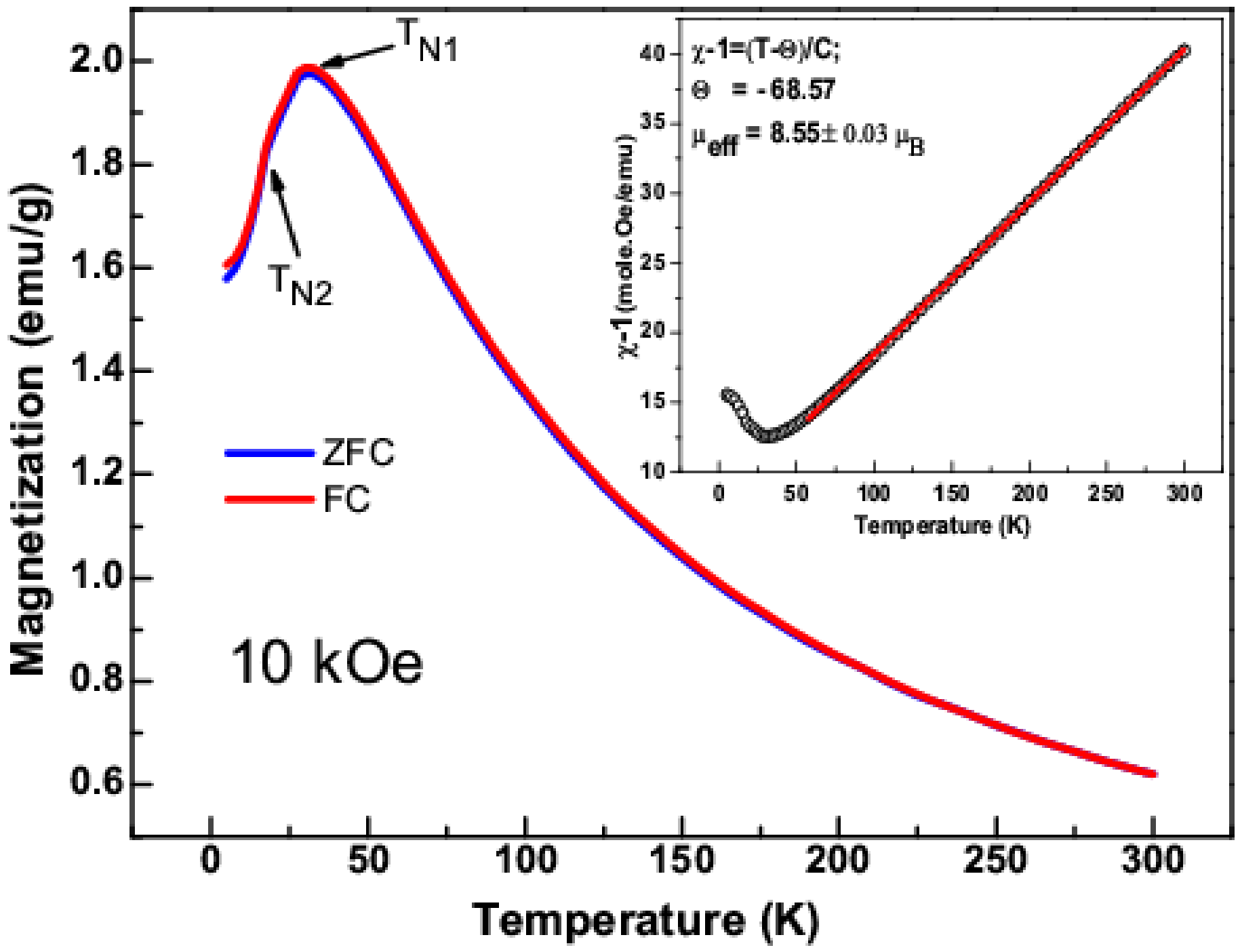